\documentstyle[12pt]{article}

\newcommand{\beq}{\begin{equation}}
\newcommand{\eeq}{\end{equation}}

\begin{document}
\def\lag{\langle}
\def\rag{\rangle}

\title{Finite Temperature Reduction\\
of the SU(2) Higgs-Model\\
with Complete Static Background\\
}

\author{{A. Jakov\'ac$^{1}$ and A. Patk\'os$^{2}$}\\
{Department of Atomic Physics}\\
{E\"otv\"os University, Budapest, Hungary}\\
}
\vfill
\footnotetext[1]{{\em e-mail address: jako@hercules.elte.hu}}
\footnotetext[2]{{\em e-mail address: patkos@ludens.elte.hu}}

\maketitle
\begin{abstract}
Direct evaluation of the 1-loop fluctuation determinant
 of non-static degrees of freedom in
a complete static background is advocated to be more efficient
for the determination of the effective
three-dimensional model of the electroweak phase transition
than the one-by-one evaluation of Feynman diagrams.
The relation of the couplings and fields of the effective model to those
of the four-dimensional finite temperature system is determined in the
general 't Hooft gauge with full implementation of renormalisation effects.
Only field renormalisation constants display dependence on the gauge fixing
parameter. Characteristics of the electroweak transition are computed from
the effective theory in Lorentz-gauge. The
dependence of various physical observables on the three-dimensional
gauge fixing parameter is investigated.
\end{abstract}

\newpage

A new wave of investigations of finite temperature gauge theories
is driven by the challenge  of the matter-antimatter asymmetry of the
Universe. Anomalous baryon number violating processes thermally excited near
the electroweak phase transition certainly have had impact on any
{\it a priori}
asymmetry. Additional non-equilibrium and CP-violating effects, occuring during
the transition, might have contributed to the generation of the present day
value of the symmetry.

Temperature introduces a natural mass-scale into the relevant field theory.
It builds up a hierarchy among the fluctuations, which should be exploited
in the evaluation of the partition function. Heavy modes with non-zero
Matsubara index are important for the accurate determination of the couplings
between the (almost) T-scale independent static modes, which drive the
phase transition. This physical picture is the content of the dimensional
reduction \cite{pisarski,landsman} of finite temperature field theories.
The validity of the assumed mass-hierarchy should be checked carefully after
each reduction step.

A correctly reduced 3-d effective model offers important advantages from the
point of view of the application of standard methods of statistical physics
to the electroweak phase transition \cite{arnold1,march,gleiser}.
Also lattice simulations are
greatly facilitated if the full 4-d system is replaced by the coresponding
3-d effective model \cite{kajantie,farakos1,karsch}, since the extreme weak
coupling situation makes
the simulation of the 4-d system a particularly involved task
\cite{bunk,montvay}.

We emphasize, that for the success of the above strategies the most
faithful possible mapping of the 4-d couplings on the temperature
dependent 3-d ones is essential. For instance, in the renormalisation
group flow of the \\
3-d model {\it dim 6} operators might play important
role. The determination of their weights in the Lagrangian of the
effective model with help of the usual Feynman diagram technique
requires calculations of increased complexity.

The first complete determination of the reduced model up to {\it dim 4}
operators in the 1-loop
approximation, including field renormalisation effects has been
published very recently \cite{farakos2}. The authors  evaluate all
relevant Feynman diagrams with two and four, zero-momentum
external field insertions. The computation has been performed in the Landau
gauge using dimensional regularisation, followed by the application of
the $\overline {{\rm MS}}$ renormalisation scheme.

In this note we present evidence that the evaluation of the functional
fluctuation determinant in a complete static background ($A_i^a({\bf
x}), A_{4}^{a}({\bf x}), \Phi ({\bf x})$) offers a simpler and more compact
calculational scheme. It allows the unified determination of all
renormalisation constants of the 4-d theory, and in principle it is easily
extendable also to the computation of the higher dimensional operators.
(After the completion of our investigation we received a paper by
Chapman \cite{chapman}, where an analogous calculation has been performed
for SU(N) pure gauge theories up to {\it dim 6} operators. The calculational
technique, however, was fully different.)

Since the method of symbolic evaluation of the functional determinant
with constant complete background is of equal difficulty for any
member of a certain gauge class, without any extra
complication one is able to study the dependence of the action of the
effective theory on the gauge fixing parameter. Specifically, we shall
perform the reduction with general 't Hooft gauge fixing, applying 3-d
momentum cut-off regularisation. The normalisation of the scalar
potential piece of the effective action will be fixed by imposing
Linde's conditions \cite{linde}.

We shall show, that the effective theory and the expressions of the 3-d
couplings do not depend on the gauge fixing parameter. The 1-loop
effective potential of the 3-d theory will be determined next in the
general (three-dimensional) Lorentz gauge, and the dependence of the
critical data ($T_c$, order parameter discontinuity, etc...) on
the parameter of the 3-d gauge fixing be discussed. This point
essentially follows \cite{arnold2}, going beyond it in the
implementation of the detailed relation between the couplings of the 3-d
theory to the 4-d ones, and the numerical evaluation of the physical
characteristics of the transition, not restricting the discussion to analytic
perturbative considerations.
\vskip 1truecm
1. The model under consideration is the SU(2) gauge+scalar theory at
finite temperature
\begin{equation}
S=\int_{0}^{\beta}d\tau \int d^{3}x\bigl [{1\over 4}F_{mn}^{a}F_{mn}^{a}+
{1\over 2}(D_{m}\Phi)^{\dagger}(D_{m}\Phi)+V(\Phi )\bigr ],
\end{equation}
\begin{equation}
V(\Phi )= {1\over 2}m^{2}\Phi^{\dagger}\Phi\
+{\lambda\over 24}(\Phi^{\dagger}\Phi)^{2},
\end{equation}
\begin{equation}
D_{m}\Phi =(\partial_{m}+igA_{m}^{a}\tau^{a}/2)\Phi,
\end{equation}
m=1,...,4; a=1,2,3. (In eqs. (1-3) the renormalised parameters appear,
the counterterms are not displayed explicitly, also Euclidean metrics is
understood). The 1-loop integration
over non-static modes will be peformed with full background, that is
all fields are split into a non-zero static and a non-static part:
\begin{eqnarray}
&
A_m=A_m({\bf x})+a_m({\bf x},\tau),\nonumber\\
&
\Phi=\left(\matrix{ 0 \cr \Phi_{0}({\bf x})\cr}\right)
          +\left(\matrix{\xi_{1}({\bf x},\tau )+i\xi_{2}({\bf x},\tau )\cr
           \xi_{3}({\bf x},\tau )+i\xi_{4}({\bf x},\tau )\cr}
\right).
\label{eq4}
\end{eqnarray}

We shall demonstrate that the full renormalised reduced action can be
recovered by choosing the static background {\it constant} (with
the most general orientation in the isospace). Upon substituting
the decomposition (4) into (1) one separates terms containing the
non-static fields up to second power, for the 1-loop integration. The
piece depending only on the constant background takes the form:
\begin{equation}
U^{(0)}=\beta V\bigl [{1\over 4}g^2(A_i\times A_j)^2+{1\over 2}
g^2(A_i\times A_4)^2+{1\over 8}g^2(A_i^2+A_4^2)\Phi_0^{\dagger}
\Phi_0+V(\Phi_0)\bigr ]
\end{equation}
(i=1,2,3). The part quadratic in the non-static components will not be
displayed explicitly, since its expression is lengthy and not
enlightening. The only important point for us is, that
the fluctuations are characterised by a $16\times 16$ matrix,
because the 12
gauge field components and 4 real Higgs scalar components are fully
coupled in the most general constant background.

The gauge fixing conditions imposed on the fluctuations $a_{m},
\xi_{\alpha} $ are
\begin{eqnarray}
&
F^1 =(D_{\mu}(A)a_{\mu})^{1}-\alpha{g\Phi_0\over 2}\xi_2,\nonumber \\ &
F^2 =(D_{\mu}(A)a_{\mu})^2-\alpha{g\Phi_0\over 2}\xi_1,\nonumber \\ &
F^3 =(D_{\mu}(A)a_{\mu})^3+\alpha{g\Phi_0\over 2}\xi_4,
\end{eqnarray}
($D_{\mu}(A)$ is the covariant derivative in the background field $A$,
$\alpha$ is the gauge fixing parameter). The corresponding
Faddeev-Popov determinant is
\begin{equation}
\det \{ [K^2 +g^2(A_i^2+A_4^2)+{\alpha \over 4}g^2\Phi_0^2 ]\delta^{a,b}-
2ig\epsilon^{abc}K_mA_m^c-g^2A_m^aA_m^b)\},
\end{equation}
where $K^2 ={\bf k}^2+\omega_n^2$.

Since the distinguishing difference of the proposed method relative to
the conventional
Feynman diagrams consists of  the explicit evaluation of the fluctuation
determinant in constant background, we are going to elaborate on
certain technical details of this computation.

The general structure of the 1-loop non-static contribution to the
reduced action looks very simple
\begin{eqnarray}
&
U^{(1)}[A_i ,A_4 ,\Phi_0 ]=
T\sum_{n\neq 0}\int{d^3k\over (2\pi)^3}\bigl\{{1\over 2}\ln [K^{32}+
\alpha^2 a_2 K^{30}+\alpha^3a_4K^{28}+...]\nonumber \\ &
{}~~~~~~~~~~~~~~~~~~~~~-\ln [K^6+...]\bigr\}
\end{eqnarray}
(The second logarithm stands for the contribution of the Faddeev-Popov
determinant).
The coefficients $a_2 ,a_4$ depend on the background fields
qudratically ($\Phi_0^2, (\hat K_m A_m)^2, A_m^2)$
  and quartically $(\Phi_0^4 , (\hat K_m A_m)^2\Phi_0^2,
(\hat K_m A_m)^4,...)$, respectively. Here $\hat K_m$ is the Euclidean
unit vector pointing in the direction of $K_m$. If one restricts the
calculation to finding the coefficients in the effective action
of all operators up to {\it dim 4}, one expands the first logarithm in
(8) up to terms $a_2^2, a_4$, and similar expansion is applied to the
contribution of the Faddeev-Popov determinant. After throwing away
(divergent) field independent terms, the coefficients we are interested
in, turn out to be proportional to various infrared safe sum-integrals
of the type
\begin{equation}
T\sum_{n\neq 0}\int{d^3k\over (2\pi)^3}\bigl( {1\over K^2}, {{\bf
k}^2\over K^4},{\omega_n^2\over K^4}, {1\over K^4},...\bigr ).
\end{equation}
In these integrals, where it is necessary, a three-dimensional
ultraviolet cut-off has been
introduced.

The complete evaluation of the fluctuation determinant with help of symbolic
programming met considerable computer memory problems.
Fortunately, in the basis, where the gauge field components are
explicitly separated
into longitudinal and transversal ones,
the diagonal elements are ${\cal O}(K^2)$, while the off-diagonals are at most
${\cal O}(K)$. Therefore, for the  first few leading
K-powers in the argument of the first logarithm of (8) it is convenient to use
a decompositon of the determinant of a certain $n\times n$ matrix $M$
into a sum of subsequent contributions containig  products of decreasing number
of diagonal elements:

$$\det M=\prod\limits_{i=1}^n M_{ii}\,\,\, + \sum\limits_{k_1>k_2}^n\!
{\rm det}_2 N(k)\!\!\!\prod\limits_{i\neq k_1,k_2}^n\!\! M_{ii} \,\,\,
+\! \sum\limits_{k_1>k_2>k_3}^n\!\!\!\!
{\rm det}_3 N(k)\!\!\!\prod\limits_{i\neq k_1\dots k_3}^n\!\!\!\! M_{ii}
\,\,\,+\dots,$$
 where
$$N(k)_{ij}=M_{k_i k_j}(1-\delta_{k_i k_j})$$
Therefore, for
$a_2$ and $a_4$ it is sufficient to consider the set of
$2\times 2$ up to $4 \times 4$ minors of the full $16\times 16$ matrix,
corresponding to the first four terms of the above decomposition.

The resulting cut-off regularised expression is the central
computational result of the present note:
\begin{eqnarray}
&
U^{(1)}[A_i, A_4, \Phi_0]=\beta V\bigl\{{1\over 2}\Phi_0^2\bigl [
({3g^2\over 16}
+{\lambda\over 12})T^2-({9g^2\over 4}+\lambda)
{\Lambda T\over 2\pi^2}\nonumber \\ &
+{m^2\over 8\pi^2}({3\alpha\over 4}g^2+\lambda)(1-D_0)\bigr ]
+{1\over 96\pi^2}\Phi_0^4(1-D_0)({27g^4\over 16}+{3\alpha g^2\lambda\over 8}
+\lambda^2)+{17g^4\over 192\pi^2}A_4^4
\nonumber \\ &
+{1\over 2}A_4^2({5g^2\over 6}T^2+{m^2g^2\over 8\pi^2}-
{5g^2\Lambda T\over 2\pi^2})
+{1\over 64\pi^2}g^2A_4^2\Phi_0^2((9-
{5\alpha \over 4})g^2+\lambda-{9+3\alpha\over 4}g^2D_0)\nonumber \\ &
{1\over 64\pi^2}g^2A_i^2\Phi_0^2({17\alpha g^2\over 12}-
{\lambda\over 3}-{9+3\alpha\over 4}g^2D_0)+{1\over 12\pi^2}g^4
(A_i\times A_4)^2({163\over 24}-\alpha-{43\over 8}D_0)\nonumber \\ &
+{1\over 24\pi^2}g^4(A_i\times A_j)^2({509\over 120}+\alpha-{43\over 8}
D_0)\bigr\}
\label{eq9}
\end{eqnarray}
with the logarithmically divergent quantity:
$D_0=\ln{\Lambda\over T}-\ln 2\pi +\gamma_{Euler}.$

It is important to call the attention of the reader to the inconvenient fact
that blindly following the
calculational scheme outlined above, two additional terms would appear
in (10):
\begin{equation}
U^{(1)}_{fake}=-\beta V({17g^4\over 960\pi^2}(A_i^2)^2+
{m^2g^2\over 48\pi^2}A_i^2).
\end{equation}
These terms clearly violate the invariance
of the final reduced theory under spatial gauge transformations. In
small periodic volumes this is the actual physical situation, since then
$A_i$ is a physical degree of freedom on the same footing as $A_4$  (spacelike
Polyakov loops are also observables). Then (11) should be added to (10)
for fields fulfilling the restriction $A_i<< 2\pi /L$ (L is the linear
spatial dimension of the system). These terms would be in complete
formal analogy with the terms representing $A_4$ in (10).

However, the expansion of the logarithms and the limit  $V\rightarrow \infty$
are not interchangeable:
\begin{equation}
\lim_{V\rightarrow\infty}\sum_{n\neq 0}\int_k\ln (1+f(A_i))\neq
\sum_{m=1}^{\infty}\lim_{V\rightarrow\infty}\sum_{n\neq 0}\int_k {(-1)^m\over
m}f(A_i)^m.
\end{equation}
This can be shown the cleanest way for "quasi-abelian" configurations
$(A_i^a=A_i\delta_{a3}, A_4=\Phi =0)$, when the left hand side of (12) is
a well-known periodic function of $A_i$ with period $2\pi/L$
\cite{weiss,gross}.
The difference between the two sides of (12) is exactly given by (11),
therefore we are led to the prescription to subtract it from the
complete result of the calculation. Only by following this careful
consideration one arrives at (10). This expression is the starting point for
the discussion of the renormalisation of the effective action.

The key observation is the renormalisation invariance of $gA_i$ in
background gauges \cite{abbott}. Exploiting this circumstance one can find all
 field renormalisations from appropriately selected terms of the sum
$U^{(0)}+U^{(1)}$. The coefficient of $g^2A_i^2\Phi_0^2/8$ determines
$Z_{\Phi}$ , that of $g^2(A_i\times A_0)^2/2$ leads to $Z_{A_0}$, and
finally $g^2(A_i\times A_j)^2/2$ to $Z_{A_i}=Z_{g}^{-1}$:
\begin{eqnarray}
&
Z_{\Phi} = 1+{17\alpha \over 192\pi^2}g^2-{1\over 48\pi^2}\lambda
-{9+3\alpha\over 64\pi^2}g^2D_0,\nonumber \\ &
Z_{A_0}=1-{163g^2\over 288\pi^2}+{\alpha g^2\over 12\pi^2}+
{43g^2\over 96\pi^2}D_0,\nonumber \\ &
Z_{A_i}=1-{509g^2\over 1440\pi^2}-{\alpha g^2\over 12\pi^2}+{43g^2\over
96\pi^2}D_0.
\end{eqnarray}
After renormalisation, these three terms are completed in view of the
minimal coupling principle into the full kinetic terms of the
corresponding fields, varying in space
\begin{eqnarray}
&
L_{3-d}^{(kin)}={1\over 4}F_{ij}^{a}F_{ij}^a+{1\over 2}(D_i(A)A_4)^2+
{1\over 2}(D_i(A)\Phi)^{\dagger}(D_i(A)\Phi),\nonumber \\ &
D_i(A){ A_4}=\partial_i A_0+g(A_i\times A_4).
\end{eqnarray}
The second step is the renormalisation of the Higgs potential. Taking
into account the effect of $Z_{\Phi}$ the regularised expression goes
over into
\begin{eqnarray}
&
U_{Higgs}^{(reg)}={1\over 2}\Phi^{\dagger} \Phi\bigl\{({3g^2\over 16}+
{\lambda\over 12})T^2-{1\over 8\pi^2}({3g^2\over 4}+\lambda)
(\Lambda^2-4\Lambda T)\nonumber \\ &
{}~~~~~~~~~~+m^2\bigl [1+{\lambda\over 6\pi^2}-{\alpha g^2\over 12\pi^2}+
{D_0\over 8\pi^2}({9g^2\over 4}-\lambda )\bigr ]\bigr\}\nonumber \\ &
{}~~~~+{1\over 24}(\Phi^{\dagger}\Phi)^2\bigl [\lambda+({27\over 8}g^4+
{\lambda^2\over 4}-{4\alpha g^2\lambda\over 3}){1\over 8\pi^2}\nonumber \\ &
{}~~~~~~~~+{D_0\over 4\pi^2}({9g^2\lambda\over 4}-{27g^4\over 16}-
\lambda^2)\bigr ].
\end{eqnarray}
It is reassuring, that the cut-off dependences of $m^2(\Lambda )$ and
$\lambda (\Lambda )$ correctly reproduce the 1-loop $\beta$-functions of
the SU(2) gauge+scalar theory (this is also true for $g^2(\Lambda )$ as
can be seen from (13)) \cite{arnold1}. For this result it is essential
to employ in the Higgs potential the correct renormalisation of the
$\Phi$-field.

The renormalisation conditions we have applied to the
tem\-pera\-ture-inde\-pendent
part of the potential, were the Linde-type
conditions, used also in our previous
publication \cite{jako}:
\begin{equation}
{dU_{Higgs}(T-indep.)\over d\Phi_0}=0,~~~~~{d^2U_{Higgs}(T-indep.)
\over d\Phi_0^2}=m_H^2(T=0),
{}~~~~\Phi_0=v_0
\end{equation}
($v_0$ is the $T=0$ expectation value of the Higgs field).
The details of the corresponding subtraction procedure were discussed in
\cite{jako} for the thermal static gauge. Here we give the final result
from the analysis done along the same lines, just for the 't Hooft gauge:
\begin{eqnarray}
&
U_{Higgs}^{(1-loop)}={1\over 2}\Phi^{\dagger}\Phi\bigl [m^2+({3g^2\over 16}
+{\lambda\over 12})T^2-{\Lambda T\over 2\pi^2}({9g^2\over 4}+\lambda )\bigr ]
+{\lambda\over 24}(\Phi^{\dagger}\Phi)^2\nonumber \\ &
{}~~~~~~~~-{1\over 64\pi^2}\sum_jn_j[m_j^4(\Phi )(\ln{m_j^2(v_0)\over T^2}+
{3\over 2})-2m_j^2(v_0)m_j^2(\Phi )],
\end{eqnarray}
where the index j runs over all formal degrees of freedom: j=4-d
transversal (T), Higgs (H), 4-d longitudinal (L), pseudo-Goldstone (G).
The corresponding quantities appearing in (17) are:
\begin{eqnarray}
&
n_T=9,~~~~~m_T^2={g^2\over 4}\Phi^{\dagger} \Phi,\nonumber \\ &
n_L=-3,~~~~~m_L^2={3g^2\over 4}\Phi^{\dagger}\Phi,\nonumber \\ &
n_H=1,~~~~~m_H^2=m^2+{\lambda\over 2}\Phi^{\dagger}\Phi,\nonumber \\ &
n_G=3,~~~~~m_G^2=-m^2+({3g^2\over 4}-{\lambda\over 6})\Phi^{\dagger}\Phi.
\end{eqnarray}
One has to emphasize that these formal degrees of freedom are not
the diagonal modes of the coupled fluctuation matrix, therefore the
masses do not correspond to any actual thermal mass. Especially, $m_G^2>0$
for the range of the $\Phi$ values between 0 and $v_0$.

The renormalisation leads also to finite rescalings of the
$A_4-\Phi$ interaction and of the $A_4$-potential, due to wave function
renormalisations:
\begin{eqnarray}
&
U^{(1-loop)}[A_4,\Phi]={17g^4\over 192\pi^2}(A_4^2)^2(1-{153g^2\over
180\pi^2}+{2\alpha g^2\over 3\pi^2})+{1\over 8}g^2A_4^2\Phi^{\dagger}\Phi
(1+{7g^2\over 10\pi^2}+{\lambda\over 6\pi^2})\nonumber \\ &
{}~~~~~~~+{1\over 2}A_4^2({5g^2\over 6}T^2+{m^2g^2\over 8\pi^2}-
{5g^2\over 2\pi^2}\Lambda T)(1-{153g^2\over 360\pi^2}+
{\alpha g^2\over 3\pi^2}).
\end{eqnarray}
The renormalised Euclidean Lagrangian density is the sum of (14),(17)
and (19). The linear divergences induced for the mass terms of $A_4$
and $\Phi$ are necessary for the mass renormalisations of the 3-d theory
at 1-loop.

It is important to note, that the effective theory shows dependence on
the gauge fixing parameter only in (19). Clearly, a 2-loop computation
of the reduced potential will also give ${\cal O}(g^4)$ contributions,
therefore the present expressions of the corrections in (19) cannot
be considered final. Omitting these
incomplete corrections, we summarize the effective theory and the
relations of the 3-d and 4-d couplings in the formulae below. It is
obvious that these relations do not involve the gauge fixing parameter.
\begin{eqnarray}
&
L_{3-d}={1\over 4}F_{ij}^aF_{ij}^a+{1\over 2}(D_i(A)A_0)^2+{1\over 2}(D_i(A)
\Phi)^{\dagger}(D_i(A)\Phi)\nonumber \\ &
+{1\over 2}m^2(T)\Phi^{\dagger}\Phi+{\hat\lambda\over 24}(\Phi^{\dagger}\Phi)^2
+{1\over 2}m_D^2A_4^2+{17g^4\over 192\pi^2}(A_4^2)^2+{\rm 3-d~ct.-terms}
\end{eqnarray}
where
\begin{eqnarray}
&
m^2(T)=\hat m^2(T)+({3g^2\over 16}+{\lambda\over 12})T^2,\nonumber \\ &
\hat m^2(T)=m^2(1-{1\over 32\pi^2}[\lambda\ln{\lambda v_0^2\over 3T^2}+
(\lambda -{9g^2\over 2})\ln{3g^2v_0^2\over 4T^2}+
{27g^4\over 4\lambda}+5\lambda-{45g^2\over 4}]),\nonumber \\ &
\hat\lambda =\lambda -{3\over 8\pi^2}[{9g^4\over 16}\ln{g^2v_0^2\over 4T^2}
+{\lambda^2\over 4}\ln{\lambda v_0^2\over 3T^2}+({\lambda^2\over 12}-
{3g^2\lambda\over 4})\ln{3g^2v_0^2\over 4T^2}
+{3\over 2} ({9g^4\over 16}+{\lambda^2\over 3}-
{3g^2\lambda\over 4})],\nonumber \\ &
m_D^2={5\over 6}g^2T^2.
\end{eqnarray}
\vskip 1truecm

2. The most adequate physical test for the effective theory seems
to be the analysis of the electroweak phase transition with help of
the effective
Higgs potential calculated from the 3-d theory in the general 3-d
Lorentz-gauge.
This has been discussed already by Arnold \cite{arnold2},
therefore we can start from his formula for the effective potential
adapted to the SU(2) case:
\begin{eqnarray}
&
U_{eff}^{(1-loop)}(\Phi_0)={1\over 2}m^2(T)\Phi_0^2+{\hat\lambda\over 24}
\Phi_0^4-{T\over 12\pi}\bigl\{6({g^2\over 4}\Phi_0^2)^{3/2}+
3({5g^2\over 6}T^2+{1\over 4}g^2\Phi_0^2)^{3/2}\nonumber \\ &
+(m^2(T)+{\hat\lambda\over 2}
\Phi_0^2)^{3/2}
+3[{1\over 2}(m_{\chi}+|m_{\chi}|
(m_{\chi}^2-\alpha g^2\Phi_0^2)^{1/2})]^{3/2}\nonumber \\ &
+3[{1\over 2}(m_{\chi}^2-|m_{\chi}|
(m_{\chi}^2-\alpha g^2\Phi_0^2)^{1/2})]^{3/2}\bigr \}.
\end{eqnarray}
In (22) the abbreviation
\begin{equation}
m_{\chi}^{2}=m(T)^2+{\hat\lambda\over 6}\Phi_0^2
\end{equation}
is introduced
and $\alpha$ is now the gauge fixing parameter of the 3-d Lorentz gauge class.

It has been argued in \cite{arnold2} that perturbative expansion of
(22) leads to a unique barrier temperature, independent of $\alpha$. We
concentrate here on $T_c$ (the transition temperature) and some further
physical characteristics of the transition, which will be determined
numerically. Also we use the detailed relationship between 3-d and 4-d
couplings, which were not taken into account in previous analyses.

$T_c$ has been determined for two characteristic values of the Higgs
mass: 60 and 80 GeV. Also the order parameter discontinuity and the surface
tension between coexisting phases have been evaluated (the latter in
the thin wall approximation). The gauge dependence of these quantities
has been tested by varying $\alpha$ in the interval (0,1). We have
restricted this interval further by requiring $U_{eff}$ to be real
in the interval of more direct physical interest,
 $\Phi\in(0,\Phi_{min}(T_c)).$ Clearly, for large enough $\alpha$
the last two terms of (22) become complex at fixed $\Phi_0$. It is
also obvious that the limiting value of $\alpha$ found in this way will
depend sensitively on the Higgs-mass $(\lambda)$ (c.f. (23)).
For $m_H(T=0)=60$ GeV
$\alpha \leq 0.3$, for $m_H(T=0)=80$ GeV $\alpha\leq 0.7$ was found to
be the "upper bound" of its allowed range of variation.

In the Table we display the relevant physical quantities, which show
remarkable stability, but definitely depend on $\alpha$. It is
interesting to note that in the same quantities calculated from an
effective potential determined in the
3-d analogue of 't Hooft's gauge, more
important variation can observed, namely of the same order of magnitude as the
difference found between 1-loop and 2-loop calculations
performed in the 4-d, finite T theory \cite{fodor,arnold3}. (For a
criticism concerning the physical interpretation of the effective potential
determined in the 't Hooft gauge, see \cite{arnold2}.)
\vskip 1truecm
3. In conclusion, we have determined in a cut-off regularized
calculation the relationship of the effective 3-d theory to those of the
original 4-d system in the general 't Hooft gauge, subject to the
renormalisation conditions (16). It has been demonstrated that the
evaluation of the determinant of non-static fluctuations in a
complete constant background is sufficient for the full specification
of the effective theory at 1-loop. The effective action proved to be
independent of the gauge fixing parameter and invariant under spatial
gauge transformations. The 1-loop analysis of the electroweak phase
transition in the effective model has been shown to be rather
insensitive to the actual choice of the gauge fixing parameter in a
general Lorentz gauge.

The method presented here is of considerable calculational advantage
over the direct enumeration  and evaluation of Feynman diagrams at
1-loop level. Prospective
 further advantages will be explored in connection of 2-loop
reduction of the standard model at high T, in the near future.
\vskip 1truecm
{\bf Acknowledgements}

The authors are grateful to Z. Fodor, K. Kajantie, M. Laine and M.
Shaposhnikov for important suggestions during the progress of this
calculation.
\vskip 1truecm
{\bf Table Caption}

Dependence of some physical observables of the electroweak phase transition
on the parameter of the 3-d Lorentz gauge fixing $\alpha$ for two typical
values of the Higgs mass ($m_H$). In subsequent columns the transition
temperature
($T_c$), the order parameter discontinuity ($\Phi_c$), the mass of the magnetic
gauge quanta $(m_W)$ at $T_c$ and the surface tension ($\sigma$) are shown.
$v_0$ is the
vacuum expectation value of the Higgs-field.

\end{document}